\documentclass[12pt,preprint]{aastex}

\shorttitle{Intrinsic Lyman-alpha Fluxes}
\shortauthors{Linsky et al.}
\slugcomment{Draft \today}

\begin{document}

\newcommand{\php}[0]{\phantom{--}}
\newcommand{\kms}[0]{km~s$^{-1}$}

\title{COMPUTING INTRINSIC LYMAN-ALPHA FLUXES OF F5 V TO M5 V STARS}

\author{Jeffrey L. Linsky\altaffilmark{1}}
\affil{JILA, University of Colorado and NIST, 440UCB Boulder, CO 80309-0440, 
USA}
\email{jlinsky@jilau1.colorado.edu}

\author{Kevin France\altaffilmark{1}}
\affil{CASA, University of Colorado, 593UCB Boulder, CO 80309-0593, USA}

\and

\author{Tom Ayres\altaffilmark{1}}
\affil{CASA, University of Colorado, 593UCB Boulder, CO 80309-0593, USA}

\altaffiltext{1}{Guest Observer, NASA/ESA {\it Hubble Space Telescope} and 
User of the Data Archive at the Space Telescope 
Science Institute. STScI is operated by the Association of Universities for 
Research in Astronomy, Inc., under NASA contract NAS 5-26555. These 
observations were made as parts of programs \#11534, \#11687, \#12034,
\#12035, and \#12464.}

\begin{abstract}

The Lyman-$\alpha$ emission line dominates the far-ultraviolet spectra 
of late-type stars
and is a major source for photodissociation of important molecules 
including H$_2$O, CH$_4$, and CO$_2$ in exoplanet atmospheres. 
The incident 
flux in this line illuminating an exoplanet's atmosphere cannot be 
measured directly as neutral hydrogen in the interstellar medium (ISM) 
attenuates most of the flux reaching the Earth. 
Reconstruction of the intrinsic 
Lyman-$\alpha$ line has been accomplished for a 
limited number of nearby stars, but is not feasible for distant or faint 
host stars. We identify correlations connecting the
intrinsic Lyman-$\alpha$ flux with the flux in other emission lines
formed in the stellar chromosphere, and find that these correlations 
depend only gradually on the flux in the other lines. These 
correlations, which are based on {\em HST} spectra, reconstructed
Lyman-$\alpha$ line fluxes, and irradiance spectra of the quiet and active 
Sun, are required for photochemical models of exoplanet 
atmospheres when intrinsic Lyman-$\alpha$ fluxes are not available. We find 
a tight correlation of the intrinsic Lyman-$\alpha$ flux with 
stellar X-ray flux for F5 V to K5 V stars, but much larger dispersion for 
M stars. We also show that knowledge of the
stellar effective temperature and rotation rate can provide   
reasonably accurate estimates of the Lyman-$\alpha$ flux for G 
and K stars, and less accurate estimates for cooler stars. 

\end{abstract}


\section{INTRODUCTION}

The discovery of large numbers of extrasolar planets (exoplanets) 
has stimulated many observational and theoretical studies of their atmospheric
chemical compositions. In particular, the habitability of an exoplanet at a 
given distance from its host star is generally thought to depend on the total 
flux received from the host star, the availability of greenhouse gasses,
and the presence of biomarkers such as O$_2$ and O$_3$ 
(but see \cite{Feng2012}). Estimates of the chemical composition
of exoplanet atmospheres depend crucially on the near-ultraviolet 
(NUV, $\lambda = $1700-3200~\AA), far-ultraviolet (FUV, 
$\lambda$ = 1170--1700~\AA), extreme ultraviolet 
(EUV, $\lambda = $300--911~\AA), XUV ($\lambda = $100-300~\AA), 
and X-ray ($\lambda < 100$~\AA) radiation from the host star, which  
control important molecular photodissociation and photoionization processes. 

Early interest concerning the influence of a star on its substellar companions
was focused on the erosion of volatile gasses from the primordial
atmospheres of the terrestrial planets of our Solar System (especially
Mars and Venus) by dissociative recombination powered by solar
ultraviolet radiation and pickup-ion stripping by 
the solar wind. These processes would 
have been enormously enhanced when the Sun was young and more magnetically
active than today \citep[e.g.,][]{Zahnle1982}.
Using a data base of solar-type stars in young clusters of known age 
and extrapolating from solar fluxes over the sunspot cycle, 
\citet{Ayres1997} showed that photoionization 
rates relevant to the Martian situation scale as power laws in time
(as anticipated in the earlier Zahnle \& Walker paper). Furthermore,
the UV flux behavior could be reliably traced back to the epoch of the Late
Heavy Bombardment
(at solar age $\sim$800~Myr), because by then rotation rates 
of young G stars had funneled into a narrow distribution from an initially
more diverse behavior \citep{Delorme2011}. 
It is the stellar spin that largely governs the 
degree of magnetic activity, and the decay of that spin by wind braking 
that causes the rapid decline of the UV and X-ray activity with increasing age.

More recently, the ``Sun in Time'' project \citep{Ribas2005,Ribas2010} 
and \citet{Sanz-Forcada2011} have
extended the earlier work to a wider range of high energy measurements,
especially including the important 912--1170~\AA\ 
{\em Far Ultraviolet Spectroscopic Explorer 
(FUSE)}\/ band, and to a broader range
of spectral types (G0 to G5), utilizing field stars whose ages must be
inferred indirectly,
but are more accessible to observation than the mostly distant cluster 
members. These studies also showed that in different wavelength bands, power
laws could be used to describe the decay of chromospheric and coronal
radiation over time.  

\citet{Ribas2005} called attention to the importance of the 
H~I Lyman-$\alpha$ line (1215.67~\AA), which is by far the brightest FUV 
emission line. They showed that for the Sun, the Lyman-$\alpha$ line flux is 
about 20\% of the total flux between 1 and 1700~\AA. The relative 
importance of the Lyman-$\alpha$ line is even more important for 
cooler stars as the photospheric emission at $\lambda > 1700$~\AA\ decreases 
rapidly with decreasing effective temperature. 

Since important molecules in planetary atmospheres, including H$_2$O, CO$_2$, 
and CH$_4$, have photodissociation cross sections that peak below 
1700~\AA\footnote{MPI-Mainz-UV-VIS Spectral Atlas of Gaseous Molecules,
http://www.atmosphere.mpg.de/enid/2295},
Lyman-$\alpha$ radiation can be the dominant cause of photodissociation
for these and other molecules. For example, the solar Lyman-$\alpha$ line
is responsible for 51\% of the photodissociation rate of H$_2$O by solar
UV radiation between 1148 and 1940~\AA, and 71\% of the total
photodissociation rate of CH$_4$ by solar radiation between 56 and 1520~\AA.
However, recent models of exoplanet atmospheres that include 
photoionization, 
e.g., models for WASP-12b \citep{Kopparapu2012}, typically do not include
realistic values for the stellar Lyman-$\alpha$ flux. In this recent model,
the FUV fluxes were obtained from the compilation of \citet{Pickles1998}
that is based on International Ultraviolet Explorer (IUE) spectra without 
a large correction for interstellar absorption in the Lyman-$\alpha$ line or 
geocoronal Lyman-$\alpha$ emission. New ultraviolet spectra of six M dwarf 
host stars observed with spectrographs on the Hubble Space Telescope
by \citet{France2012b} provide an observational basis for more realistic 
photochemical models of exoplanet atmospheres.

While exoplanet atmospheres absorb the 
Lyman-$\alpha$ flux from their host stars without attenuation, 
neutral hydrogen in the interstellar medium scatters most of the Lyman-$\alpha$
flux out of the line-of-sight between the star and Earth. 
One must therefore correct for interstellar absorption by reconstructing the 
stellar Lyman-$\alpha$ profile and flux. For pre-main sequence stars with 
disks, \citet{Herczeg2004} and \citet{Schindhelm2012} showed that the 
fluorescent H$_2$ emission lines pumped by Lyman-$\alpha$ provide a useful 
diagnostic for reconstructing the intrinsic Lyman-$\alpha$ emission line.
For 62 main-sequence and giant stars without H$_2$ fluorescence, 
\citet{Wood2005} reconstructed 
Lyman-$\alpha$ fluxes using interstellar H~I column densities and velocities
inferred from the deuterium Lyman-$\alpha$ and metal lines. 
\citet{France2012a} developed an alternative reconstruction technique
in which the widths and strengths of one or two Gaussian emission lines 
representing Lyman-$\alpha$ and 
the velocities and column densities of interstellar
absorption features are iteratively varied to obtain a solution that
best fits the observed Lyman-$\alpha$ line wings.
Both techniques require high-resolution spectra with good signal/noise
and minimal contamination by geocorona Lyman-$\alpha$ emission. The only
available spectrograph that can obtain such data is the Space Telescope 
Imaging Spectrograph (STIS) on {\em HST}.
High instrumental sensitivity, especially important for faint M dwarfs, 
increasing interstellar hydrogen column densities with distance, 
and difficult-to-obtain {\em HST} observing time limit the number of 
stars for which Lyman-$\alpha$ fluxes can be reconstructed by these two 
techniques.  

The objective of this study is to find different methods for infering 
the Lyman-$\alpha$ flux incident on the atmospheres of exoplanets for stars 
when high-resolution Lyman-$\alpha$ spectra are not available. 
These methods should be applicable to a broad range of stars including
M dwarfs, which are very numerous and could support habitable planets with
short orbital periods. One proposed method is to use correlations
of reconstructed Lyman-$\alpha$ line fluxes with other stellar
features (lines of O~I and C~II) formed in the same temperature range as 
Lyman-$\alpha$, lines of Mg~II and Ca~II formed at slightly lower 
temperatures, and lines of C~IV formed at somewhat higher temperatures. Our
hypothesis is that ratios of the Lyman-$\alpha$ line flux to fluxes in
these other lines should be similar for stars in limited ranges of spectral 
type with only a small dependence on stellar activity as indicated by 
the emission line fluxes. Empirical support for this hypothesis comes from 
the similar power law slopes of UV emission lines \citep{Ayres1997,Ribas2005}
and the correlation of Lyman-$\alpha$ and Mg~II fluxes \citep{Wood2005}. 
The similar shapes of the chromosphere and transition region thermal structures
for regions on the Sun with very different heating rates
\citep{Fontenla2011} provides theoretical support for our hypothesis.

The paper is structured as follows. In section 2 we describe our selection 
of targets that have reconstructed Lyman-$\alpha$ and other UV emission line 
fluxes. 
Section 3 describes our procedure for reconstructing the 
intrinsic Lyman-$\alpha$ flux using correlations with other UV emission lines
and the errors and uncertainties of this method. Section 4 describes an 
extension of this procedure using correlations with ground-based Ca~II H and K
line fluxes, and section 5 describes how correlations with X-ray fluxes can be 
used to predict the intrinsic Lyman-$\alpha$ flux. Section 6 describes how 
the stellar effective temperature and rotation rate
can provide rough estimates of the intrinsic Lyman-$\alpha$ flux when no 
emission line fluxes are available. The final section lists our conclusions. 
In a subsequent paper, we will apply the flux correlation technique to 
estimating stellar EUV radiation, which is responsible for photoionizing 
hydrogen and other species, heating outer atmospheres, and driving mass 
loss from exoplanets.

\section{TARGET SELECTION}

Our target list consists of all F5 to M5 main sequence stars for which 
reconstructed Lyman-$\alpha$ fluxes are now available. 
\citet{Wood2005} reconstructed Lyman-$\alpha$ line profiles observed by STIS
in its medium (spectral resolution $\approx \lambda/45,000$) and 
high-resolution ($\approx \lambda/100,000$) echelle modes using 
interstellar deuterium Lyman-$\alpha$ and metal absorption 
lines to model the interstellar H~I column density as a function of wavelength 
across the hydrogen Lyman-$\alpha$ line. They provided reconstructed 
Lyman-$\alpha$ line fluxes for 40 main-sequence stars with spectral types
F5 to M5, together with Mg II h and k line and X-ray fluxes for 
most of these stars. We also include five M dwarf stars that are known to 
host exoplanets. 
The Lyman-$\alpha$ fluxes for these stars were reconstructed from STIS 
grating and echelle spectra using an iterative  technique to measure the
intrinsic Lyman-$\alpha$ profile
and the interstellar absorption \citep{France2012a}.

For most these stars, we have measured fluxes of the 
H~I 1215.67~\AA, O~I 1304.86~\AA\ + 1306.03~\AA,
C~II 1335.71~\AA, and C~IV 1548.19~\AA\ + 1550.77~\AA\ lines. 
In the semi-empirical solar chromosphere and transition region model  
computed by \citet{Avrett2008}, 
the peak contributions to the line center emission for these lines are at 
temperatures near 40,000~K, 6,700~K, 29,500~K, and 68,000~K, respectively.
However, the wings of these lines are formed over a range of cooler 
temperatures. We did not use
fluxes of the O~I 1302.17~\AA\ and C~II 1334.53~\AA\ lines as 
these lines show strong interstellar absorption. We extracted the 
O~I, C~II and C~IV line fluxes from Cosmic Origin Spectrograph 
(COS) \citep{Green2012} data with spectral resolution 
($\approx \lambda/17,000$) available through the Mikulski Archive for 
Space Telescopes (MAST)\footnote{http://archive.stsci.edu} and
the well-calibrated STIS spectra from the StarCAT \citep{Ayres2010}
website\footnote{casa.colorado.edu/$\sim$ayres/StarCAT/}. 
In a few cases we used data obtained with the Goddard High Resolution 
Spectrograph (GHRS) instrument on HST. The O~I data obtained with COS 
are not usable because of airglow contamination. We subtracted the 
underlying continuum from the emission line fluxes. Line fluxes for the
solar-mass stars were measured by \citet{Linsky2012}.

We also include solar spectral irradiance data observed with the 
Solar Radiation and Climate Experiment (SORCE) 
on the Solar-Stellar Irradiance Comparison 
Experiment II (SOLSTICE II) \citep{Woods2009,Snow2005}. These are 
integrated sunlight spectra directly comparable with the stellar data. 
The 2008 April 10--16 data refer to the very quiet Sun, and the 2003 October
27 and November 10 data refer to the Sun when it was very active.

Table 1 summarizes the stellar and emission line fluxes f(line) in 
ergs cm$^{-2}$ s$^{-1}$ at a distance of 1 AU for 45 stars, the quiet Sun, 
and the active Sun at two different times. Since most of the stars are 
variable, especially in the FUV, the fluxes refer to a single time and
are not time averages. For most of the stars the STIS Lyman-$\alpha$
spectra and the COS spectra of the O~I, C~II, and C~IV were obtained at 
different times. Since stellar UV radiation varies with time especially 
for M dwarfs and active warmer stars,
ratios of the Lyman-$\alpha$ flux to other
emission lines will have systematic errors compared to the stars for which
all of the emission lines were observed at nearly the same time.

\section{CORRELATIONS OF RECONSTRUCTED LYMAN-$\alpha$ FLUX WITH EMISSION LINES 
FORMED IN THE CHROMOSPHERE AND TRANSITION REGION}

\subsection{F5 V to G9 V stars}

Table 1 lists the measured fluxes for 21 F5--G9 stars and the Sun at
three different times and activity levels. We plot in Figure 1 the flux ratio 
R(C~IV) = f(Lyman-$\alpha$)/f(C~IV) vs. f(C IV) with line fluxes measured in 
ergs~cm$^{-2}$~s$^{-1}$ at a distance of 1~AU. Figures 2--5 show similar 
plots of R(line) = f(Lyman-$\alpha$)/f(line) vs. f(line) using the 
previously described C~II, O~I, and Mg~II lines. The R(line) 
ratios for the F5--G9 stars in Figures 1--4
follow tight trajectories of decreasing R(line) with increasing f(line). 
In Table~1 we list the iron abundances relative to hydrogen 
[Fe/H] listed in the SIMBAD\footnote{http://simbad.u-strasbg.fr/simbad} data
base. Three of the F5--G9 stars (HR~4657, $\chi$~Her, and $\tau$~Cet) 
have large iron depletions (defined as [Fe/H] $<0.3$), and often have higher
R(line) values than other stars with similar fluxes
and [Fe/H] values close to the solar. This behavior is expected since R(line)
should increase as the line's metal abundance decreases, but whether or not 
R(line) is inversely proportional to [Fe/H] will be 
considered in the next section. We therefore exclude the low [Fe/H]
stars when computing least-squares fits for the remaining F5--G9 stars. 
The coefficients for these fits,
log[R(line)] = A + Blog[f(line)], are listed in Table 2. 
We note that the least-squares fits, which do not include the solar data,
are close to the solar data. 
Also, the $\alpha$~Cen~A ratios are similar to the quiet and active
Sun ratios even though the solar and stellar data are measured by different 
instruments. Table 2 shows that the mean dispersions of the F5--G9 stars 
about the best fit lines are 18--24\%.

\subsection {K0 V to K5 V stars}

There are a total of 16 stars in this group: 15 have Mg~II fluxes, and 8 have 
C~IV, C~II fluxes and O~I fluxes. The least-squares fits to
the R(line) ratios for the K0~V -- K5~V stars have similar slopes to 
the fits to the R(line) ratios for the F5--G9 stars and similar dispersions
(15--30\%) about the fit lines.  
The K0 V-K5 V stars have a wide range in rotational periods 
(0.38 days for Speedy Mic to 43 days for $\alpha$~Cen~B), 
but there is no strong trend of R(line) with rotational period and thus with
activity. Eight of these stars are classified in SIMBAD as either BY~Dra type 
varables or RS~CVn type variables, but we can find no information on the 
companion stars which are likely fainter and cooler than the primary stars. 
  
\subsection{M0 V to M5 V stars}

We analyze spectra of nine M0~V to M5.5~V stars with reconstructed 
Lyman-$\alpha$ fluxes. Five of these are exoplanet host stars. 
We plot R(line) vs. 
log[f(line)] in Figures 1--3 and 5. The dispersion of the data about the 
fit lines is much larger for the M stars than for the F5--K5 stars
primarily due to their much larger time variability as described in the 
following section. The dispersion in R(O I) for the M stars, however, may be 
unrepresentative given the small number of data points.

\subsection{Errors and Uncertainties}

In this analysis, we identify four causes of uncertainty in the R(line) values:
flux measurement errors, Lyman-$\alpha$ line reconstruction errors, 
uncertain atomic abundances, and unknown time variability. 
We can estimate the error magnitudes 
or at least identify
the first three types of errors, but time variability errors cannot be 
measured with the existing data and are therefore systematic. 

The smallest errors are typically flux-measurement errors for the C~IV, C~II,
O~I, and Mg~II lines. These are typically less than 2\% for the STIS data 
in StarCAT and the COS data sets, but can be as large as 5--20\% for the 
faintest M dwarfs. Uncertainties in the reconstruction of the 
Lyman-$\alpha$ lines are usually larger than the line flux measurement 
errors. \citet{Wood2005} estimated 20\% for typical errors in reconstructing
Lyman-$\alpha$ line fluxes. In order to compare the 
two reconstruction methods, 
we re-fit the observed Lyman-$\alpha$ line profiles of
several stars in the \citet{Wood2005} sample with the 
iterative least-squares technique \citep{France2012a}. For the favorable
case of AU Mic, a star with high signal/noise data and a single ISM absorption 
feature, the agreement in the reconstructed fluxes between the two techniques 
is within 5\%. For the unfavorable case of AD Leo, a star with  
several ISM velocity components along the line of sight, 
the disagreement is about 30\%. In Figures 1-5, we plot representative errors
of 20\% for the R(line) ratios based on the \citet{Wood2005} 
reconstruction, or 30\% for the faint M dwarfs based on the 
\citet{France2012a} 
reconstruction. As shown Table 2, the mean dispersions of the F5--G9 and 
K0--K5 stars (ignoring the low metal abundance stars) are in the range 
18--30\%, which is consistent with our estimates of the Lyman-$\alpha$ 
reconstruction errors and smaller time variability errors.   

We have excluded stars with low iron abundances ([Fe/H]$<0.3$) from the sample 
of stars used to make the least-squares fits. 
The [Fe/H] abundances for the F, G, and K stars were obtained from the SIMBAD
data base.
Accurate abundance measurements for M dwarfs require special techniques given 
the complexity of their spectra. Recent work includes calibration of
the offset of stars in the H-R diagram relative to the main sequence  
\citep{Johnson2009} and
measurements of Na~I and Ca~I absorption lines in near-IR spectra
\citep{Rojas2010}. \citet{Rojas2012} have revised their earlier work 
and compared their new results to earlier studies. 
We therefore adopt their [Fe/H] measurements for AD~Leo, EV~Lac, and the 
exoplanet host stars GJ~876, GJ~581, and GJ~436. We adopt
the [Fe/H] value for GJ~832 measured by \citep{Rojas2010}. 
\citet{Cayrel2001} lists
[Fe/H]=--0.54 for GJ~667C, but we do not expect this star to have a 
significant underabundance as the other host stars show normal or superrich
abundances. The young star AU Mic should have near solar abundances, and
Proxima~Cen should have the near-solar abundances of $\alpha$~Cen A and B.

For each low iron abundance star, Table~4 lists the value of [Fe/H], 
and for each spectral line the log of the difference, which we refer to as 
``delta'', between R(line) and the least-squares fit line at the value of 
f(line). If the deltas were only due to low
metal abundance and the abundances of carbon, oxygen, and magnesium relative 
to the Sun were the same as for iron, then the observed deltas  
should be the negative of the log iron depletions. This
is not the case, but the data in Table~4 show some interesting trends.
For example, the deltas seen in the C~IV and C~II lines 
(and to some extent in the O~I line) are nearly the same for each star, 
implying that the abundances of carbon and oxygen may be important 
factors in determining the deltas. However, the deltas are much 
smaller than --[Fe/H] for the F5--G9 and K0--K5 stars.
Thus metal depletion cannot be the only factor determining the deltas, 
and the thermal structures of the atmospheres of stars of different 
spectral type likely play a role. Since the many 
lines of Fe, Mg, and Ca are the main cooling agents of the lower solar 
chromosphere \citep{Anderson1989}, lower abundances of these elements
will change the energy balance and thus the thermal structure.

Since the deltas for the Mg~II lines of the F5--K5 stars are mostly smaller 
than those for 
the C~IV, C~II, and O~I lines, there will likely be a smaller uncertainty 
when estimating R(Mg~II) from the least-squares fits compared to using R(line) 
for the C~IV, C~II, and O~I lines. 
However, it is important to correct for 
interstellar absorption in the Mg~II lines, which requires high-resolution 
spectra. Since interstellar absorption is minimal for the C~IV lines, the C~II
1335~\AA\ line, and the O~I 1304 and 1306~\AA\ lines, lower resolution spectra
of these lines can be used without interstellar corrections for estimating 
R(line) and thus the intrinsic Lyman-$\alpha$ flux. 

In most cases, the STIS spectra containing the Lyman-$\alpha$ line and the 
COS or STIS spectra used for extracting the other line fluxes 
were obtained at different times. 
For these stars the R(line) ratios will depend on time variations in the 
stellar UV emission. 
The importance of time variability for different types of stars can be
estimated from the mean dispersion and RMS dispersions of the R(line)
ratios about the least-squares fits. These dispersions for the M stars listed 
in Table~2 are much larger than for the F5--K5 stars and much larger than
the 20--30\% uncertainty associated with the Lyman-$\alpha$ reconstructions. 
This behavior is expected since 
M stars are more variable and flare more often than the warmer stars.

\section{CORRELATION OF RECONSTRUCTED LYMAN-$\alpha$ FLUX WITH Ca II EMISSION}

The hydrogen H$\alpha$ (6563~\AA) and Ca~II H and K (3933 and 3968~\AA) lines 
formed in the chromosphere are observable by ground-based telescopes. 
However, the H$\alpha$ line is difficult to analyze since 
increasing heating first deepens the absorption line and then produces an 
emission feature. We consider instead the Ca~II lines because the emission 
in the centers of the
broad photospheric absorption lines is a good indicator of the chromospheric 
heating rate. An important difficulty is distinguishing the chromospheric 
emission inside of the H1 and K1 features that define the emission core
from the photospheric emission that would be present in the 
absence of a chromosphere. For active stars and especially 
M dwarfs, this is not a large uncertainty as the photospheric emission is
faint compared to the bright chromospheric emission. For the less active 
G stars, the photospheric emission is comparable to or larger than the 
chromospheric emission. It is therefore essential to correct for the 
photospheric emission as we wish
to correlate chromospheric Ca~II emission with the intrinsic Lyman-$\alpha$ 
flux.

Different authors has estimated the photospheric emission in and near the
core of the Ca~II lines in different ways. Using high-resolution spectra, 
\citet{Linsky1979} and \citet{Pasquini1988} measured the flux in the Ca~II 
emission line cores between the H1 and K1 minima features and subtracted 
the small amount of flux in these wavelength intervals predicted by 
radiative equilibrium model photospheres. 
\citet{Robinson1990} and \citet{Browning2010} fitted the photospheric
H and K absorption lines with Gaussians and then subtracted the flux 
interpolated in the line cores from the measured Ca~II emission. This approach 
somewhat overcorrects for the photospheric emission, but the error is small for
the very cool stars that they observed. The resulting
chromospheric Ca~II fluxes at 1~AU are listed in columns 6 and 7 of Table~3.

Other observers have measured the S index, which is the flux in 1.09~\AA\ 
bandpasses centered on the H and K line emission cores divided by the flux 
in continuum windows. 
\citet{Hartmann1984} provided a correction for the photospheric flux
within the bandpass but outside of the H1 and K1 features using 
high-resolution spectra of stars with weak H and K emission. Columns 8 and 9
of Table~3 show the time-averaged Ca~II fluxes obtained at Mt. Wilson 
by \citet{Noyes1984} and \citet{Baliunas1995} after subtracting the 
photospheric flux following \citet{Hartmann1984}. Columns 10 and 11 show Ca~II
fluxes obtained by \citet{Henry1996}, \citet{Hall2007,Hall2009}, and 
\citet{Lockwood2007} using a similar approach. The last column in Table~3
lists the fluxes of Southern hemisphere stars obtained by 
\citet{Cincunegui2007}. We plot in Figure~6 the ratio of Lyman-$\alpha$
to Ca~II flux, R(Ca~II), using the average of the Ca~II fluxes for all of the 
data in Table~3, except for the \citet{Cincunegui2007} fluxes which appear 
to be systematically smaller than the fluxes obtained from the other sources..

The R(Ca~II) data in Figure~6 for each spectral type bin show relatively small
scatter about the least-squares fit lines. We find that the dispersion 
(see Table~2) is smallest for the K stars (only 20.6\%), and larger for the 
variable M stars. The dispersion of 42\% for the G stars 
likely results from uncertainty in 
the photospheric flux correction which is large for G stars. Thus the Ca~II
H and K line flux can be used to estimate the intrinsic Lyman-$\alpha$ flux  
from the fit lines, provided one can remove the photospheric component from
the observed line core fluxes. 

\section{CORRELATION OF RECONSTRUCTED LYMAN-$\alpha$ FLUX WITH X-RAY EMISSION}

Except for the highly variable M dwarfs for which the Lyman-$\alpha$ and other
emission lines were observed at different times, the correlations of 
reconstructed Lyman-$\alpha$ flux with emission lines formed at similar or 
somewhat higher temperatures in the chromosphere and transition region 
generally have small scatter about the least-squares fit lines.
This behavior results from increasing mechanical heating shifting the 
chromospheric thermal structure to higher densities while keeping a similar 
shape \citep{Fontenla2011}. Thus all of the emission lines brighten together 
roughly proportional to density squared.
\cite{Ayres1997}, \cite{Ribas2005}, and others have noted that 
slopes of power-law relations between X-ray and chromosphere line flux 
are generally near 2.0 rather than a linear relation. Also, soft X-ray emission
as measured, for example, by ROSAT is highly sensitive to coronal temperature 
and is more time variable than chromospheric emission. For these
reasons, we did not anticipate a tight correlation 
between reconstructed Lyman-$\alpha$ flux and X-ray flux. \cite{Wood2005}
plotted X-ray flux vs reconstructed Lyman-$\alpha$ flux for F--M dwarfs and 
giants showing correlations, but with large scatter for each type of star.

We plot in Figure~7 the R(X)=f(Lyman-$\alpha$)/f(X-ray) ratios vs. X-ray flux 
at 1~AU for the stars in Table~1 using the ROSAT PSPC X-ray fluxes cited by
\cite{Wood2005} or more recent X-ray flux measurements using Chandra and 
XMM-Newton (see the references cited in the footnotes of Table~1). 
The X-ray flux for GJ~667C, which was obtained with the HRI instrument on 
ROSAT rather than the PSPC, is uncertain due to the close proximetry of 
GJ~667A and GJ~667B. \citet{Judge2003} estimated solar X-ray fluxes through
the ROSAT PSPC bandpass from full disk solar observations by the 
SNOE-SXP instrument. For the quiet Sun, moderately active Sun, and active Sun,
we use their X-ray luminosities, $\log L_X =$ 26.8, 27.75, and 27.9, 
respectively.   

The Lyman-$\alpha$/X-ray flux data show correlations with small scatter for 
the F5--G9 and K0--K5 dwarfs. 
The quiet and active solar data are consistent with the
least-squares fit for the other F5--G9 stars. 
The mean dispersions listed in Table~2 are 29.8\% for the 
F5--G9 dwarfs and 22.2\% for the K0--K5 dwarfs, which are similar to 
the mean dispersions for the C~IV, C~II, O~I, Mg~II, and Ca~II lines. 
The mean dispersion for the M dwarfs is
much larger than for the F5--K5 stars, although the slope of the fit line
is similar to that of the warmer stars. X-ray time variability is the most
likely cause of the large scatter of the M stars as the X-ray 
and Lyman-$\alpha$ data were obtained at different times. Despite the factor
of 2--3 scatter about the M star fit line, the similar slope of the fit line 
to those for the warmer stars
suggests that one can use the M dwarf fit line to estimate the Lyman-$\alpha$ 
flux of an M dwarf at the time of an X-ray measurement with an uncetainty much 
smaller than the dispersion shown in Table~2. We note that GJ832 is plotted 
very close to the quiet Sun in Figure 7, and that GJ832 (M1.5~V) and 
AU~Mic (M0~V) are consistent with the fit lines for the K0--K5 stars.
This suggests that the R(X) vs. X-ray flux correlation for K stars may also 
be useful for estimating the intrinsic Lyman-$\alpha$ flux of early M dwarfs.

\section{CORRELATION OF THE RECONSTRUCTED LYMAN-$\alpha$ FLUX WITH 
STELLAR EFFECTIVE TEMPERATURE AND ROTATION}

We now consider whether the stellar Lyman-$\alpha$ flux of main-sequence stars 
can be estimated simply from the stellar effective temperature 
($T_{\rm eff}$). We plot in Figure~8
the Lyman-$\alpha$ flux for all of the stars in Table~1 vs. $T_{\rm eff}$.
At each value of $T_{\rm eff}$, there is a dispersion in the  Lyman-$\alpha$
flux that increases from a factor of 4 near $T_{\rm eff}=6000$~K to a
factor of 1000 for the M stars. The range in Lyman-$\alpha$ flux at a given
$T_{\rm eff}$ is reduced significantly by grouping the stars according to their
rotation period ($P_{\rm rot}$), which is a rough measure of magnetic heating 
rates in the chromosphere and corona. We show least squares fits, 
log f(Lyman-$\alpha$) = A + BT$_{\rm eff}$, for the stars 
in three groups: fast rotators ($P_{\rm rot}$ = 3--10 days),
moderate rotators (10--25 days), and slow rotators ($>25$ days). 
The coefficients for these fits are listed in Table~5. Compared to the 
least-squares fits, the Lyman-$\alpha$ flux dispersion within each 
rotational period group is 32--85\% and largest for the M stars. Therefore,
estimates of the Lyman-$\alpha$ flux based only on a star's $T_{\rm eff}$
and $P_{\rm rot}$ are reasonably accurate for G stars but not for 
the cooler stars.

In Figure~9, we plot the Lyman-$\alpha$ flux in the habitable zone (HZ) of an 
exoplanet, where the distance from the star to the habitable zone is
estimated approximently as proportional to the stellar luminosity,
$d_{\rm HZ}=(R_{\star}/R_{\odot})(T_{\star}/T_{\odot})^2$ AU.
The location of the HZ, the region around a star where life forms are thought 
to be possible, involves many other considerations including greenhouse gases,
orbital eccentricity and stability, and atmospheric chemical composition
\citep{Kasting2003}.
In this HZ plot, the dispersion of the Lyman-$\alpha$ flux is
reduced somewhat for the M stars (see Table~5). 
Lyman-$\alpha$ fluxes in the HZ trend significantly higher
with decreasing $T_{\rm eff}$, and the quiet Sun has nearly the smallest
value of Lyman-$\alpha$ flux. 
Exoplanets of most of the stars discussed in this 
paper receive significantly larger Lyman-$\alpha$ fluxes and thus have higher
molecular dissociation rates in their atmospheres. This is especially true
for M dwarfs, which have smaller radii and lower effective temperatures 
than the Sun. Exoplanets in the HZs of the five M dwarfs in this study 
receive about 10 times the Lyman-$\alpha$ flux as the Earth receives from 
the quiet Sun. We therefore conclude that large FUV fluxes are received 
by exoplanets in their HZs from their M dwarf host stars contrary to 
some previous assumptions.
Also, the Lyman-$\alpha$ fluxes for stars with exoplanets do not appear
to differ from stars without discovered exoplanets in these two plots.

\section{CONCLUSIONS}

We have identified five methods for reconstructing or estimating the 
intrinsic Lyman-$\alpha$ 
stellar fluxes of main-sequence stars between spectral types F5 and M5. 
Whether a given method can be used and its accuracy depends on the 
quality of the available data. The first and most accurate method
developed by \citet{Wood2005} requires high-resolution spectra of the 
Lyman-$\alpha$ line and knowledge of the interstellar velocity structure
based on high-resolution spectra of the deuterium Lyman-$\alpha$ and
metal lines. \citet{Wood2005} estimates typical errors of 20\% in the 
reconstructed Lyman-$\alpha$ line fluxes using this method, but at present
only the STIS instrument on HST can provide such data for nearby stars.

The second method developed by \citet{France2012a} also
requires high-resolution spectra of the hydrogen Lyman-$\alpha$ line, 
but does not require spectra of the deuterium Lyman-$\alpha$ line or any 
other interstellar absorption line. This technique
solves for both the intrinsic Lyman-$\alpha$ line parameters and the 
interstellar absorption simulataneously. When the interstellar absorption 
has only one velocity component, the technique can be as accurate as the first,
but it fails when the interstellar velocity structure has many components,
which is not known in the absence of high-resolution interstellar absorption 
lines.

In this paper, we considered three additional methods that can be used
with different accuracy when there is no available high-resolution spectrum 
of the Lyman-$\alpha$ line to serve as the basis for reconstruction. 
The third method described in Sections 3 and 4 requires flux measurements 
of the stellar C~IV, C~II, O~I, Mg~II, or Ca~II lines and the best fit 
correlations of these lines with fluxes of the reconstructed Lyman-$\alpha$ 
lines in our data set. 
This method estimates Lyman-$\alpha$ fluxes with 18--25\% uncertainty 
for F5--K5 dwarf stars, provided one corrects high-resolution spectra 
of the Mg~II lines for interstellar absorption. 
This method is based on the hypothesis that the ratios of the Lyman-$\alpha$ 
line flux to C~IV, C~II, O~I, Mg~II, and Ca~II line fluxes, R(line), 
for stars of similar spectral type depend only gradually on line flux. 
We find that this hypothesis is valid for late-type dwarf stars with 
approximently solar abundances.
Using this method for F5--K5 V stars, we find that the dispersions of R(C~IV),
R(C~II), R(O~I), and R(Mg~II) about the least-squares fits are 
consistent with the likely errors in reconstructing the  
intrinsic Lyman-$\alpha$ fluxes. The dispersion of R(Ca~II) about the fit line
for the F5--G9 stars is larger than for the other lines because of 
uncertain estimates of the photospheric flux.

Most dispersions for the M0--M5~V stars are significantly larger, 
probably due to stellar variability between the time the 
Lyman-$\alpha$ line was observed with the STIS instrument and the other lines
were observed with COS or STIS with a different grating setting. Even with 
this time variability, the fit lines for the O~I, C~II, and C~IV lines
should provide estimates of the intrinsic Lyman-$\alpha$ flux for a given star
with an uncertainty less than a factor of two. We suggest that estimating the 
intrinsic Lyman-$\alpha$ flux from the Mg~II line flux and the 
least-squares fit will lead to smaller 
uncertainty provided one has high-resolution Mg~II spectra with which to 
estimate interstellar absorption in these lines.

It is important, however, to recognize the limitations of this correlation 
method.
The errors associated with measuring fluxes of the Mg~II, O~I, C~II, and 
C~IV lines are generally small at the time of these line flux measurements, 
although geocoronal emission through the 
large COS aperture can be important for the O~I lines. 
Since the method does not use direct measurements of the Lyman-$\alpha$ line, 
there is no reconstruction error.
In addition, there is no time variability error as the Lyman-$\alpha$
flux is the scaled value at the time the other emission lines are 
observed. The largest uncertainty is the stellar metal abundance. 
Stars known to have low metal abundances generally have 
R(line) ratios above the scaling relation predictions. Since the effect 
appears to be smallest for the Mg~II lines, 
correlations with the Mg~II line fluxes may 
provide more accurate intrinsic Lyman-$\alpha$ line fluxes. 
Observations of more stars,
especially M stars, are need to better understand the errors in R(line) 
associated with low metal abundances.

When no UV or Ca~II line fluxes are available but there are X-ray measurements 
with an energy range similar to that of the ROSAT PSPC, a fourth method can 
provide estimates of the intrinsic Lyman-$\alpha$ flux from least-squares fits
to the intrinsic Lyman-$\alpha$/X-ray flux ratio vs. X-ray flux. The
mean dispersions about the fit lines for F5--G9 dwarfs and K0--K5 dwarfs
are 20--30\%, but the mean dispersion for the M dwarfs is much larger
again due to the large time variability of X-ray emission and the 
comparison of X-ray and Lyman-$\alpha$ data obtained at different times.
Since the slope of the Lyman-$\alpha$/X-ray flux ratio vs. X-ray flux 
for the M stars is similar to that obtained for the warmer stars, we suggest 
that one can obtain a factor of two estimate of the intrinsic Lyman-$\alpha$ 
line flux at the time of the X-ray measurement from the correlation fit line.

Even when no Lyman-$\alpha$, UV emission lines, or X-ray data are available, 
one can use a fifth method to estimate the intrinsic Lyman-$\alpha$ flux 
for F5--M5 stars 
based only on the star's effective temperature and some measure of 
stellar activity. The plot of intrinsic Lyman-$\alpha$ flux vs. stellar
effective temperature shows more than an order of magnitude range in
Lyman-$\alpha$ fluxes at a given effective temperature, but comparing 
data for stars of similar rotational period, a good measure of stellar 
activity, significant reduces the range. Comparison of Lyman-$\alpha$
fluxes vs. effective temperature for stars of similar rotational period 
as viewed from their habitable zones further reduces the mean dispersion about 
the fit lines to 30--40\%. Thus even in the absence of any spectroscopic data, 
this fifth method can provide useful estimates of the intrinsic
Lyman-$\alpha$ flux of F5--M5 dwarf stars.

\acknowledgements

This work is supported by NASA through grants NNX08AC146, NAS5-98043, 
and HST-GO-11687.01-A to the University of Colorado at Boulder. 
We thank Martin Snow for providing the SORCE data, and Jurgen Schmitt for
providing X-ray luminosities for M dwarf stars. We thank the 
anonymous referee for
his thorough critique of the initial manuscript. JLL thanks the 
Kiepenheuer-Institut f\"ur Sonnenphysik in Freiburg Germany for hospitality
while writing this paper.

{\it Facilities:} \facility{HST (COS)}, \facility{HST (STIS)}. 
\facility{SIMBAD}

\begin{deluxetable}{lccccccccccc}
\tablewidth{0pt}
\tabletypesize{\scriptsize}
\rotate
\tablenum{1}
\tablecaption{Stellar Line Fluxes (ergs cm$^{-2}$ s$^{-1}$) at 1 Astronomical 
Unit}
\label{tab:log}
\tablehead{\colhead{Star\tablenotemark{a}} & 
\colhead{HD} & 
\colhead{[Fe/H]} &
\colhead{Spec Type\tablenotemark{b}} & 
\colhead{d(pc)\tablenotemark{b}} &
\colhead{Lyman-$\alpha$\tablenotemark{c}} &
\colhead{Mg II h+k\tablenotemark{d}} & \colhead{O I 130.4+130.6} & 
\colhead{C II 133.6} & \colhead{C IV 154.8+150.1} &
\colhead{X-ray} & 
\colhead{Ref}\tablenotemark{e}}
\startdata
Procyon      & 61421   &--0.02 & F5 IV-V&3.50 & 77.1 & 267  & 1.77  & 2.58  &
4.62  & 11.5 &5\\
HR 4657      & 106516  &--0.70 & F5 V/L & 22.6 & 27.8 & 62.6 & 0.418:& 0.320:&
0.493:& 5.43 &1\\
$\zeta$ Dor  & 33262   &--0.15 & F7 V & 11.7 & 46.5 & 108  & 0.627 & 1.050 &
1.958& 16.7 &2\\
$\chi$ Her   & 142373  &--0.50 & F8 V/L & 15.9 & 22.0 & 45.2 & 0.235 & 0.115 &
0.171 & 0.312 &1\\
--           & 28033   &+0.11  &  F8 V & 46.4 & 24.7 & 50.4 &  --   &  --   &
  -- & 19.2 &--\\
$\chi^1$ Ori & 39587   &--0.09 &  G0 V & 8.66 & 41.6 & 79.8 & 0.474 & 0.684 &
1.247 & 37.3 &2\\ 
HR 4345      & 97334   &--0.01 &  G0 V & 21.93& 42.8 & 108  & 0.569 & 1.016 &
1.631 & 39.9 &2\\
SAO 136111   & 73350   &+0.07  & G0 V & 23.98& 32.8 & 65.8 & 0.296 & 0.429 &
0.737 & 19.3 &2\\
V993 Tau     & 28205   &+0.12  & G0 V & 47.01& 55.5 & 140  & 0.901 & 1.286 &
2.46: & 41.4 &1\\
V376 Peg*    & 209458  &--0.06 & G0 V & 49.63& 15.7 & --   &  --   &  --   & 
--  & -- &--\\

Quiet Sun (4/2008)&    &+0.00  & G2 V & --    & 5.95 & 18.2 & 0.0793& 0.0862&
0.129 & 0.224 &3\\
Active Sun (11/2003)&  &+0.00  & G2 V & --   & 7.04 & --   & 0.0881& 0.1041&
0.152 & 1.99 &3\\
Active Sun (10/2003)&  &+0.00  & G2 V & --   & 9.15 & --   & 0.1107& 0.1497&
0.212 & 2.85 &3\\
$\alpha$ Cen A & 128620&+0.25  & G2 V & 1.325& 7.54 & 29.7 & 0.1032& 0.1057&
0.161 & 0.117 &4,13 \\
HR 2882      & 59967   &--0.19 & G4 V & 21.82& 55.9 & 90.2 & 0.550 & 0.821 &
1.881 & 41.9 &2 \\
61 Vir*      & 115617  &+0.00  & G5 V & 8.555& 5.26 & 14.1 & 0.0488&0.0427 &
0.0603 & 0.265 &1\\
$\kappa^1$ Cet & 20630 &+0.09  & G5 V & 9.14 & 30.0 & 53.0 & 0.366 & 0.556 &
0.873 & 25.6 &2\\
HR 2225      & 43162   &+0.00  & G5 V & 16.72& 41.0 & --   & 0.396 & 0.479 &
0.835 & 48.1 &1 \\
HR 6748      & 165185  &--0.06 & G5 V & 17.55& 48.9 &101.1 & 0.495 & 0.687 &
1.123 & 53.5 &2 \\
SAO 254993   & 203244  &--0.21 & G5 V & 20.42& 43.8 & 59.1 & 0.311 & 0.539 &
0.940 & 20.2 &1\\
SAO 158720   & 128987  &+0.01  & G6 V & 23.68& 34.4 & 60.1 & 0.296 & 0.381 &
0.510 & 14.3 &1\\
$\tau$ Cet   & 10700   &--0.43 & G8 V/L& 3.65& 5.66 & 7.93 & 0.0417& 0.0229&
0.0327 & 0.176 &5\\
$\xi$ Boo A  & 131156A &--0.13 & G8 V & 6.70 & 35.3 & 61.9 & 0.274 & 0.465 &
0.815 & 28.3 &1\\ 
SAO 28753    & 116956  &+0.03  & G9 IV-V&21.9 & 33.0 & 57.5 & 0.336 & 0.504 &
0.781 & 24.7 &1\\
\hline

$\alpha$ Cen B*&128621 &+0.24  & K0 V & 1.255& 10.1 & 19.1 & 0.0809& 0.0925& 
0.132 & 0.533 &5,13\\
DX Leo       & 82443   &--0.23 & K0 V & 17.8 & 31.1 & 79.1 & --    & --    &
--  & 59.7 &--\\
70 Oph A     & 105341  &--0.08 & K0 V & 5.09 & 23.6 & 30.6 & 0.222 & 0.296 & 
0.431 & 6.62 &5\\
HR 8         & 166     &+0.15  & K0 V & 13.67& 37.9 & 57.0 & 0.374 & 0.592 & 
1.036 & 33.0 &1\\
$\epsilon$ Eri*&22049  &--0.08 & K1 V & 3.216& 21.5 & 27.2 & 0.145 & 0.179 & 
0.274 & 5.63 &1,9\\
40 Eri A     & 26965   &--0.27 & K1 V & 4.985& 7.33 & 12.3 & --    & --   &  
--  & 1.15 &--\\
36 Oph A     & 155886  &--0.39 & K1 V/L& 5.464& 18.0& 14.0 & 0.0816& 0.0873&
--     & 3.72 &--\\
HR 1925      & 37394   &+0.14  & K1 V & 12.28& 29.3 & 45.0 & 0.228 & 0.277 & 
0.462 & 14.4 &1\\
--*           & 189733  &--     & K1 V & 19.25& 11.8 & --   & 0.202 & 0.274 & 
--   & 5.34 &5,9\\
EP Eri        & 17925   &+0.08  & K2 V & 10.35& 27.6 & 61.1 & --    &  --   & 
0.917 & 32.9 &5\\
LQ Hya       & 82558   &--     & K2 V &18.62 & 59.1 & 75.2 & --    &   --  & 
4.744 & 243. &5\\
V368 Cep     & 220140  &--     & K2 V & 19.20& 46.9 & 86.6 & --    &  --   &
--    & 275. &--\\
PW And       & 1405    &--     & K2 V & 21.9 & 47.1 & 51.8 & --    &  --   & 
3.154 & 187. &5\\
Speedy Mic   & 197890  &--1.49 & K2 V/L& 52.2& 214  & 190  & --    & --    &
--   & 4255. &--\\
61 Cyg A     & 201091  &--0.35 & K5 V/L& 3.487& 8.90& 7.35 & 0.0353& 0.0323& 
--   & 0.498 &5,13\\
$\epsilon$ Ind &209100 &--0.20 & K5 V & 3.622& 17.3 & 8.43 & --    & --    &
--   & 0.871 &--\\
\hline
AU Mic       & 197481  &--     & M0 V & 9.91 & 43.0 & 17.6 & 0.360 & 0.525 & 
1.035 & 70.9 &1,8\\
AD Leo       & GJ 388  &+0.28  & M3.5 V&4.695& 9.33 & 2.19 & 0.0644& 0.137 & 
0.376 & 19.1 &1,10\\
EV Lac       & GJ 873A &--0.01 & M3.5 V&5.122& 3.07 & --   & 0.0163& 0.0402& 
0.1094 & 19.5 &1,10\\
Proxima Cen  & GJ 551C &--     & M5.5 V&1.296& 0.301& --  &0.000408&0.00162&
0.00719 & 0.142 &1,11\\
GJ 832*      & 204961  &--0.12 & M1.5 V  &4.954&5.17 & 0.311& --   &0.00278&
0.00762 & 0.214 &6,9\\
GJ 667C*     & --      &--     &M1.5 V   &6.9  &1.54 & 0.113& --   & --    &
--     & 0.263 &12\\
GJ 876*      & --      &+0.18  & M5.0 V&4.689& 0.409&0.0315& --    &0.00874&
0.0186 & 0.0574 &7,9\\
GJ 581*      & --      &--0.10 &M2.5 V  &6.3 & 0.513&0.0360&  --   &0.000820&
0.00304 & -- &6\\
GJ 436*      & --      &+0.04  &M3 V  &10.3& 1.571 &0.1038 & --    &0.00182 &
0.00623 & 0.0334 &6,9\\

\enddata
\tablenotetext{a}{Exoplanet host stars listed in the exoplanets.org data base
are marked with a * symbol.} 
\tablenotetext{b}{Data from SIMBAD with M star metal abundances from the 
sources listed in Section 3.4.}. 
\tablenotetext{c}{Reconstructed intrinsic Lyman-$\alpha$ line fluxes for most  
stars \citep{Wood2005} and for the GJ stars \citep{France2012b} .}
\tablenotetext{d}{Mg II h and k core emission corrected for interstellar 
absorption for most stars \citep{Wood2005} and for GJ stars (this paper).}
\tablenotetext{e}{Sources of the O I, C II, and C IV data: (1) StarCAT
\citep{Ayres2010}, (2) \citet{Linsky2012}, (3) Martin Snow (private 
communication), (4) \citet{Pagano2004}, (5) This paper, 
(6) \citet{France2012a}, (7) \citet{France2012b}, 
(8) X-ray flux in \citet{Schneider2010},
(9) X-ray flux in \citet{Sanz-Forcada2011}, 
(10) X-ray flux in \citet{Robrade2005}, 
(11) X-ray flux in \citet{Guedel2004}, 
(12) X-ray flux from J. Schmitt (private communication)
(13) X-ray flux in \citet{Robrade2012}.} 

\end{deluxetable}

\begin{deluxetable}{lccccccc}
\tablewidth{0pt}
\tabletypesize{\scriptsize}
\rotate
\tablenum{2}
\tablecaption{Least-squares Fits to Line Flux Ratios}
\label{tab:log}
\tablehead{\colhead{Star} & \colhead{Number} & 
\colhead{Spectral} & 
\colhead{Flux\tablenotemark{a}} &
\colhead{A\tablenotemark{b}} &
\colhead{B\tablenotemark{b}} & \colhead{Mean} & 
\colhead{RMS}\\ 
\colhead{Group} & \colhead{Included} &
\colhead{Line} & \colhead{Range} & \colhead{} & \colhead{} &
\colhead{Dispersion (\%)} & \colhead{Dispersion (\%)}}

\startdata
F5 V -- G9 V & 16 & C IV & 0.06 to 4.62 & 1.560 & --0.353 & 18.4 & 22.2\\  
K0 V -- K5 V & 8  & C IV & 0.13 to 4.74 & 1.508 & --0.575 & 15.6 & 18.0 \\
M0 V -- M5 V & 8  & C IV & 0.003 to 1.04& 1.325 & --0.355 & 118.5 & 159.3\\

F5 V -- G9 V & 16 & C II & 0.043 to 2.58& 1.723 & --0.298 & 19.4 & 22.1\\
K0 V -- K5 V & 6  & C II & 0.09 to 0.59 & 1.731 & --0.314 & 25.0 & 35.6\\
M0 V -- M5 V & 8  & C II & 0.0008 to 0.53& 1.525 & --0.404& 113.2 & 156.4\\

F5 V -- G9 V & 16 & O I  & 0.049 to 1.77& 1.872 & --0.198 & 22.5 & 27.2\\
K0 V -- K5 V & 6  & O I  & 0.08 to 0.37 & 1.886 & --0.198 & 22.3 & 32.8\\
M0 V -- M5 V & 4  & O I  & 0.0004 to 0.36&1.871 & --0.278 & 16.9 & 17.8\\

F5 V -- G9 V & 16 & Mg II& 14.1 to 267  &--0.291& --0.0208 & 24.2 & 31.2\\
K0 V -- K5 V & 12 & Mg II& 8.4 to 86.6  & 0.338& --0.318  & 29.8& 39.3\\
M0 V -- M5 V & 7  & Mg II& 0.032 to 17.6& 0.814 & --0.296 & 27.8 & 33.8\\

F5 V -- G9 V & 13 & Ca II& 11.4 to 149. & --0.223 & 0.080 & 42.2 & 49.6\\
K0 V -- K5 V & 10 & Ca II& 5.9 to 39.4  & 0.605 & --0.433 & 20.6 & 30.7\\
M0 V -- M5 V & 7  & Ca II& 0.0076 to 6.53&1.028 & --0.312 & 39.9 & 49.6\\

F5 V -- G9 V & 20 & X-ray& 0.12 to 52.5 & 1.156 & --0.684 & 29.8 & 42.3\\
K0 V -- K5 V & 15 & X-ray& 0.50 to 3090 & 1.058 & --0.707 & 22.2 & 27.4\\
M0 V -- M5 V &  8 & X-ray& 0.033 to 70.9& 0.431 & --0.573 & 122. & 155.\\
\enddata
\tablenotetext{a}{Flux range of the C~IV, C~II, O~I, and Mg~II lines and 
X-ray flux in ergs cm$^{-2}$ s$^{-1}$ at a distance of 1 AU.}
\tablenotetext{b}{lLeast-squares fit to 
log[f(Lyman-$\alpha$)/f(line)] = A + B log[f(line)],
where the line is C~IV, C~II, O~I, or Mg~II.}
\end{deluxetable}

\begin{deluxetable}{lcccccc|ccccc}
\tablewidth{0pt}
\tabletypesize{\scriptsize}
\rotate
\tablenum{3}
\tablecaption{Ca II H and K  Line Fluxes (ergs cm$^{-2}$ s$^{-1}$) at 
1 Astronomical Unit}
\label{tab:log}
\tablehead{\colhead{Star\tablenotemark{a}} & 
\colhead{HD} & 
\colhead{[Fe/H]} &
\colhead{Spec Type\tablenotemark{b}} & 
\colhead{Lyman-$\alpha$\tablenotemark{c}} &
\colhead{Ca II\tablenotemark{d}} & 
\colhead{Ca II\tablenotemark{e}} &
\colhead{Ca II\tablenotemark{f}} &
\colhead{Ca II\tablenotemark{g}} &
\colhead{Ca II\tablenotemark{h}} &
\colhead{Ca II\tablenotemark{i}} &
\colhead{Ca II\tablenotemark{j}}}
\startdata
Procyon      & 61421   &--0.02 & F5 IV-V & 77.1 & & & & 149. & & & \\
HR 4657      & 106516  &--0.70 & F5 V/L & 27.8 & & & 46.4 & 40.1 & & & \\
$\chi$ Her   & 142373  &--0.50 & F8 V/L & 22.0 & & & 31.5 & 34.9 & & 25.5 &\\
$\chi^1$ Ori & 39587   &--0.09 &  G0 V & 41.6 & & & 48.3 & 53.0 & & 52.7 &\\ 
HR 4345      & 97334   &--0.01 &  G0 V & 42.8 & & & 49.6 & 54.5 & & 52.0 &\\

Mean Sun      &    &+0.00  & G2 V & 6.5~ & 10.8 & & & 16.1 & & 17.2 &\\

$\alpha$ Cen A & 128620&+0.25  & G2 V & 7.54 & 16.6 & & & & & 15.8 & 7.75\\
HR 2882      & 59967   &--0.19 & G4 V & 55.9 & & & & & 44.7 & & 27.4\\
61 Vir*      & 115617  &+0.00  & G5 V & 5.26 & & & & 11.3 & 11.6 & &\\
$\kappa^1$ Cet & 20630 &+0.09  & G5 V & 30.0 & 64.1 & & 39.4 & 45.2 & & 43.0 &\\
HR 6748      & 165185  &--0.06 & G5 V & 48.9 & & & & & 45.6 & & 21.8\\
SAO 254993   & 203244  &--0.21 & G5 V & 43.8 & & & & & 33.0 & & 23.8\\
$\tau$ Cet   & 10700   &--0.43 & G8 V/L& 5.66 & & & 6.61 & 7.24 & 6.52 & 6.46 & 4.61\\
$\xi$ Boo A  & 131156A &--0.13 & G8 V & 35.3 & 31.5 & & 23.9 & 26.9 & 27.1 & 23.3 &\\ 
\hline

$\alpha$ Cen B*&128621 &+0.24  & K0 V & 10.1 & 5.75 & & & & 6.02 & & 3.05\\
DX Leo       & 82443   &--0.23 & K0 V & 31.1 & & & & 39.4 & & &\\
70 Oph A     & 105341  &--0.08 & K0 V & 23.6 & 23.1 & & & & & &\\
$\epsilon$ Eri*&22049  &--0.08 & K1 V & 21.5 & 19.7 & & 15.9 & 16.8 & 14.8 & 15.9 & 9.37\\
40 Eri A     & 26965   &--0.27 & K1 V & 7.33 & 7.99 & & 7.42 & 8.08 & 6.41 & & 3.70\\
36 Oph A     & 155886  &--0.39 & K1 V/L& 18.0 & & & 10.4 & 10.5 & 9.39 & &\\
HR 1925      & 37394   &+0.14  & K1 V & 29.3 & & & & 23.4 & & &\\
EP Eri        & 17925   &+0.08  & K2 V & 27.6 & & & 34.2 & 34.1 & 32.9 & & 22.2
\\
61 Cyg A     & 201091  &--0.35 & K5 V/L& 8.90 & 1.61 & & 3.0 & 3.28 & 7.89 & 
3.29 &\\
$\epsilon$ Ind &209100 &--0.20 & K5 V & 17.3 & 3.29 & & & & 8.13 & & 2.92\\
\hline

AU Mic       & 197481  &--     & M0 V & 43.0 & 8.60 & 4.46 & & & & &\\
AD Leo       & GJ 388  &+0.28  & M3.5 V& 9.33 & & 0.264 & & & & & 0.958\\
EV Lac       & GJ 873A &--0.01 & M3.5 V& 3.07 & & 0.314 & & & & &\\
Proxima Cen  & GJ 551C &--     & M5.5 V& 0.301 & & & & & & & 0.0067\\
GJ 667C*     & --      &--     &M1.5 V &1.54 & & 0.103 & & & & &\\
GJ 876*      & --      &+0.18  & M5.0 V& 0.409 & & 0.00765 & & & & &\\
GJ 581*      & --      &--0.10 &M2.5 V & 0.513 & & 0.00828 & & & & &\\
GJ 436*      & --      &+0.04  &M3 V & 1.571 & & 0.0975 & & & & &\\

\enddata
\tablenotetext{a}{Exoplanet host stars listed in the exoplanets.org data base
are marked with a * symbol.} 
\tablenotetext{b}{Data from SIMBAD with M star metal abundances from the 
sources listed in Section 3.4.}. 
\tablenotetext{c}{Reconstructed intrinsic Lyman-$\alpha$ line fluxes for most  
stars \citep{Wood2005} and for the GJ stars \citep{France2012b} .}
\tablenotetext{d}{Ca II H and K chromospheric flux at 1 AU 
\citep{Linsky1979}, \citet{Pasquini1988} and \citep{Robinson1990}.} 
\tablenotetext{e}{Ca II H and K chromospheric flux at 1 AU 
\citep{Browning2010}.} 
\tablenotetext{f}{Ca II H and K chromospheric flux at 1 AU 
\citep{Noyes1984}.} 
\tablenotetext{g}{Ca II H and K chromospheric flux at 1 AU 
\citep{Baliunas1995}.} 
\tablenotetext{h}{Ca II H and K chromospheric flux at 1 AU 
\citep{Henry1996}.} 
\tablenotetext{i}{Ca II H and K chromospheric flux at 1 AU 
\citep{Hall2007,Hall2009,Lockwood2007}.} 
\tablenotetext{j}{Ca II H and K chromospheric flux at 1 AU 
\citep{Cincunegui2007}.} 

\end{deluxetable}

\begin{deluxetable}{lcccccc}
\tablewidth{0pt}
\tablenum{4}
\tablecaption{Iron Depletions and Differences between R(line) and 
Least-Squares Fits}
\tablehead{\colhead{Star} & \colhead{Spectral Type} & 
\colhead{[Fe/H]} & \multicolumn{4}{c}{Difference in dex of R(line) relative 
to Fits}\\
\colhead{} & \colhead{} & \colhead{} & 
\colhead{C IV} & \colhead{C II} & \colhead{O I} & \colhead{Mg II}}
\startdata
HR 4657    & F5 V & --0.70  & 0.08 & 0.07 & --0.12 & --0.02\\
$\chi$~Her & F8 V & --0.50  & 0.28 & 0.28 & --0.02 & 0.01\\
$\tau$ Cet & G8 V & --0.43  & 0.15 & 0.18 & --0.01 & 0.16\\
36 Oph A   & K1 V & --0.39  & --   & 0.25 & 0.24   & 0.14\\
Speedy Mic & K2 V & --1.49  & --   & --   & --     & 0.44\\
61 Cyg A   & K5 V & --0.35  & --   & 0.24 & 0.25   & 0.02\\
\enddata
\end{deluxetable}

\begin{deluxetable}{lccccc}
\tablewidth{0pt}
\tablenum{5}
\tablecaption{Least-Squares Fits to Lyman-$\alpha$ Flux vs. Effective 
Temperature and Rotation Rate}
\tablehead{\colhead{Distance} & \colhead{$P_{\rm rot}$} & 
\colhead{A} & \colhead{B} & \colhead{Mean} & \colhead{RMS}\\ 
\colhead{from Star} & \colhead{(days)} & \colhead{} & \colhead{} &
\colhead{Dispersion(\%)} & \colhead{Dispersion(\%)}}
\startdata
1 AU  & 3--10  & 0.37688 & 0.0002061    & 41.0 & 73.6\\
1 AU  & 10--25 & 0.48243 & 0.0001632    & 32.3 & 42.1\\
1 AU  & $>25$  & --1.5963& 0.0004732    & 85.0 & 99.8\\
HZ    & 3--10  & 3.9358  & --0.0004054  & 34.8 & 46.7\\
HZ    & 10--25 & 4.5460  & --0.0005631  & 37.4 & 44.0\\
HZ    & $>25$  & 3.5737  & --0.0004686  & 43.1 & 51.7\\
\enddata
\end{deluxetable}

\begin{figure}
\includegraphics{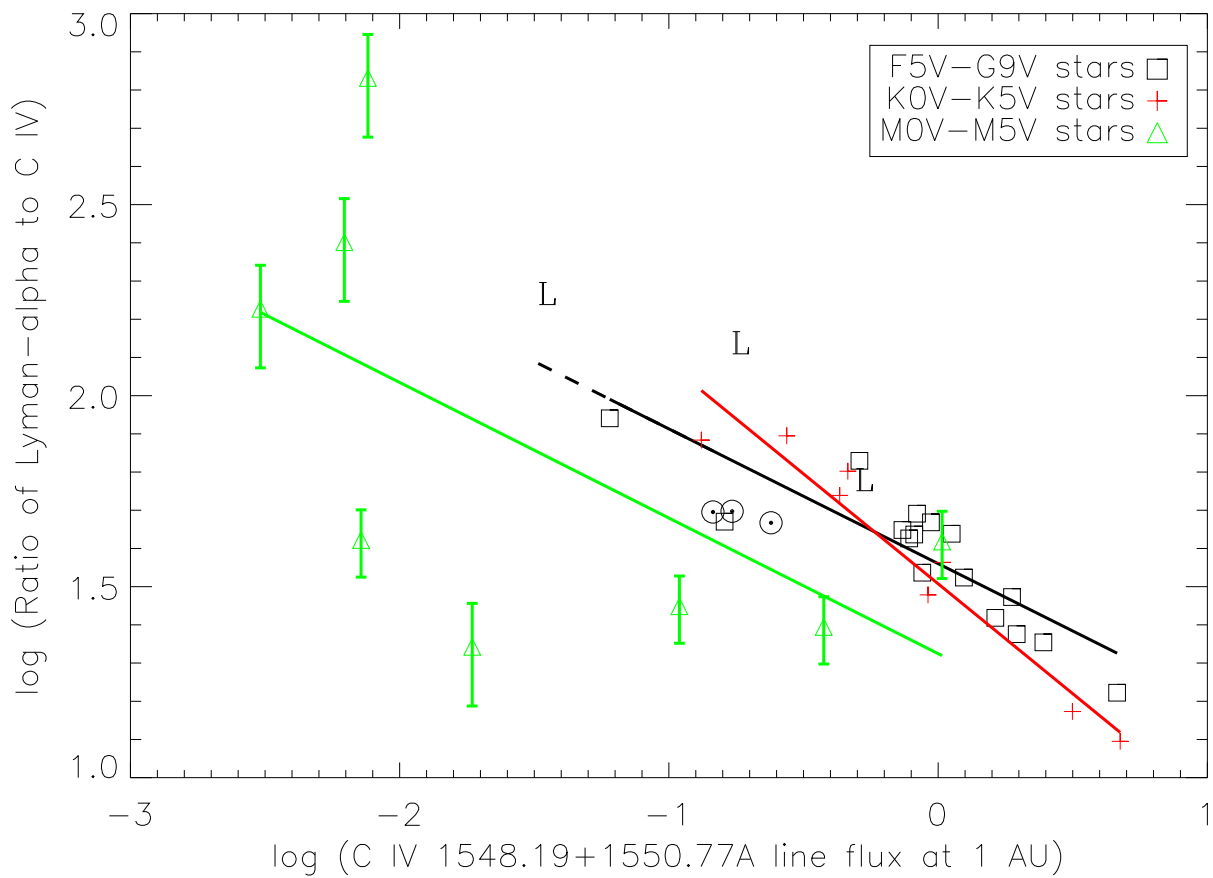}
\caption{Plot of the ratio of the Lyman-$\alpha$ to C IV 1548+1550~\AA\ line 
flux vs. the C IV line flux at 1 AU. Included are stars between spectral types 
F5 V and M5 V, divided into three spectral type bins, the quiet Sun and the 
active Sun at two different times. The solar data are indicated by Sun symbols,
and the L symbol refers to a star with low metal abundance [Fe/H] $<-0.30$. 
The solid lines are least-squares fits for each spectral type bin
excluding the L stars and the Sun.
The ratio for $\alpha$~Cen~A is closest to the solar ratios. The errors bars 
are 20\% for stars using the
\citet{Wood2005} correction for missing Lyman-$\alpha$ flux or 30\% for stars
using the \citet{France2012a} correction for missing Lyman-$\alpha$ flux.}
\end{figure} 

\begin{figure}
\includegraphics{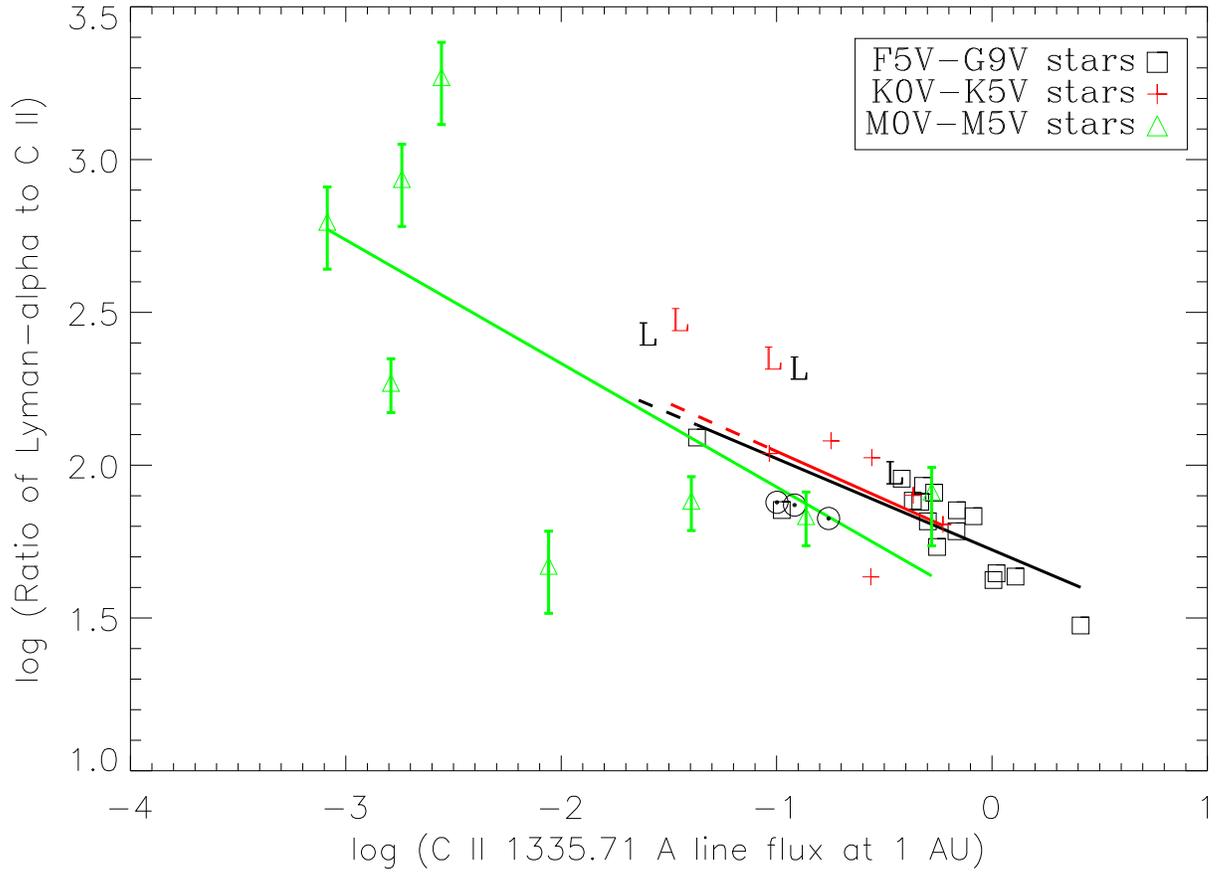}
\caption{Same as Figure 1 except for the Lyman-$\alpha$ to C~II 1335~\AA\ 
line flux ratio vs. the C~II line flux at 1 AU.}
\end{figure} 

\begin{figure}
\includegraphics{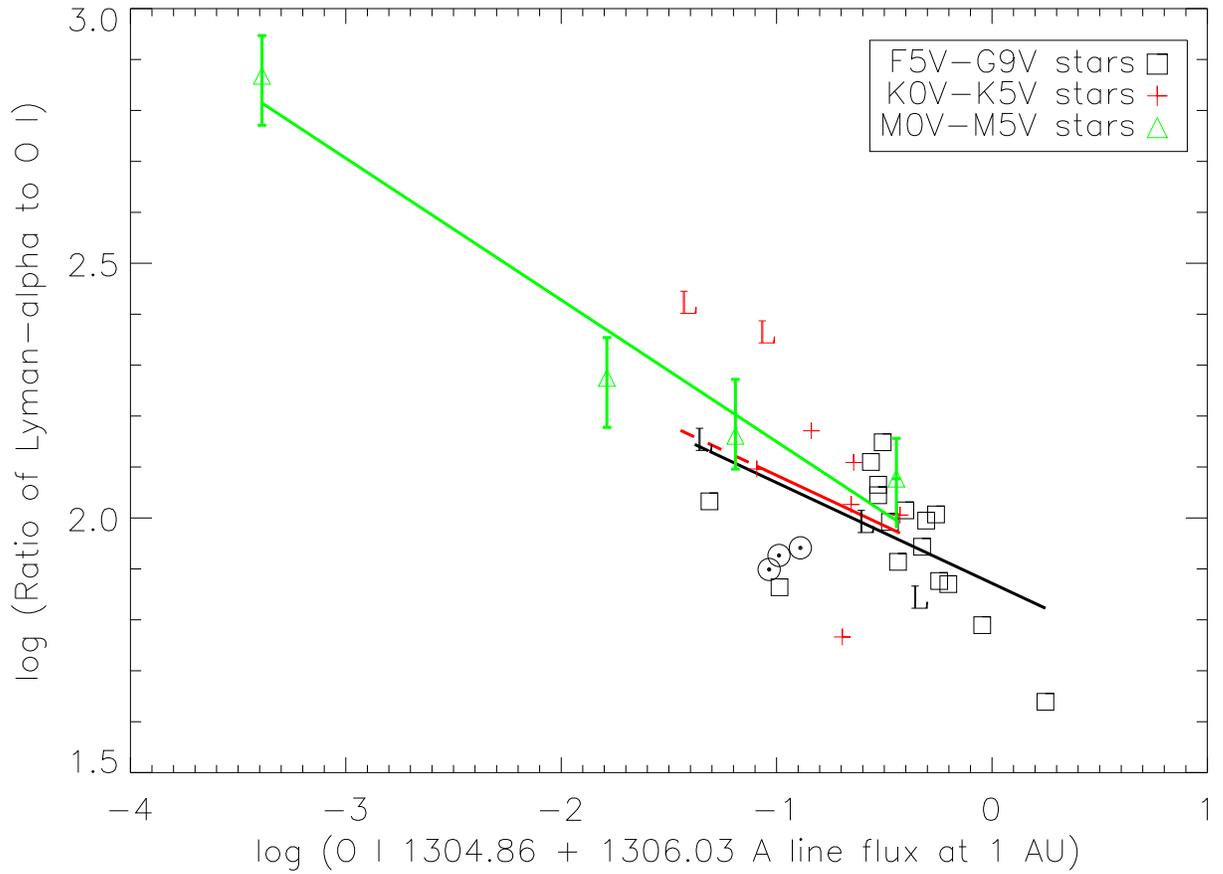}
\caption{Same as Figure 1 except for the Lyman-$\alpha$ to O~I 1304.86 + 
1336.03~\AA\ line flux ratio vs. the O~I line flux at 1 AU. The small 
dispersion of the line flux ratios for the M stars may be unrepresentative 
given the small number of data points.}
\end{figure} 

\begin{figure}
\includegraphics{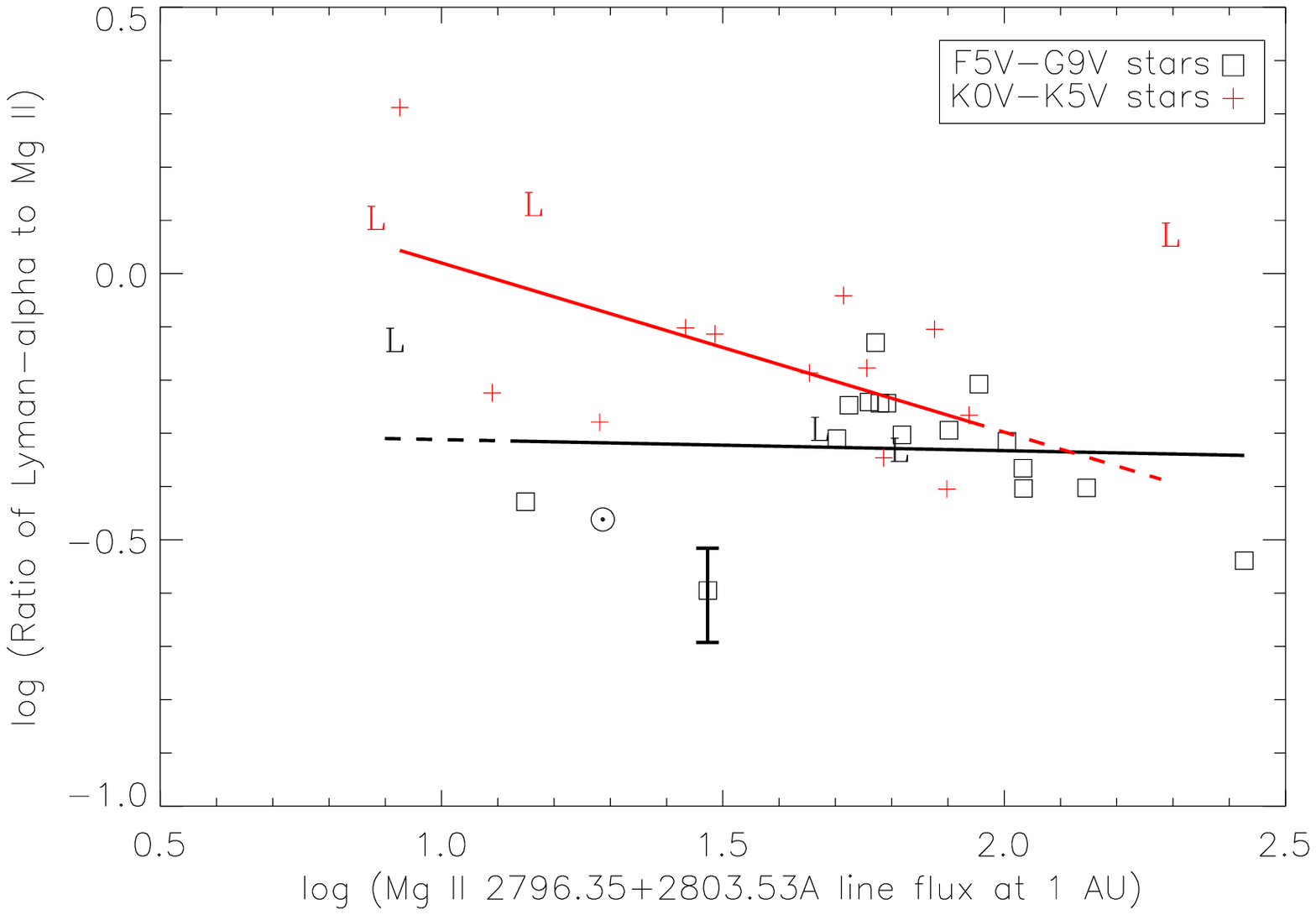}
\caption{Same as Figure 1 except for the Lyman-$\alpha$ to Mg~II 2796 + 
2803~\AA\ line flux ratio vs. the Mg~II line flux at 1 AU.}
\end{figure} 

\begin{figure}
\includegraphics{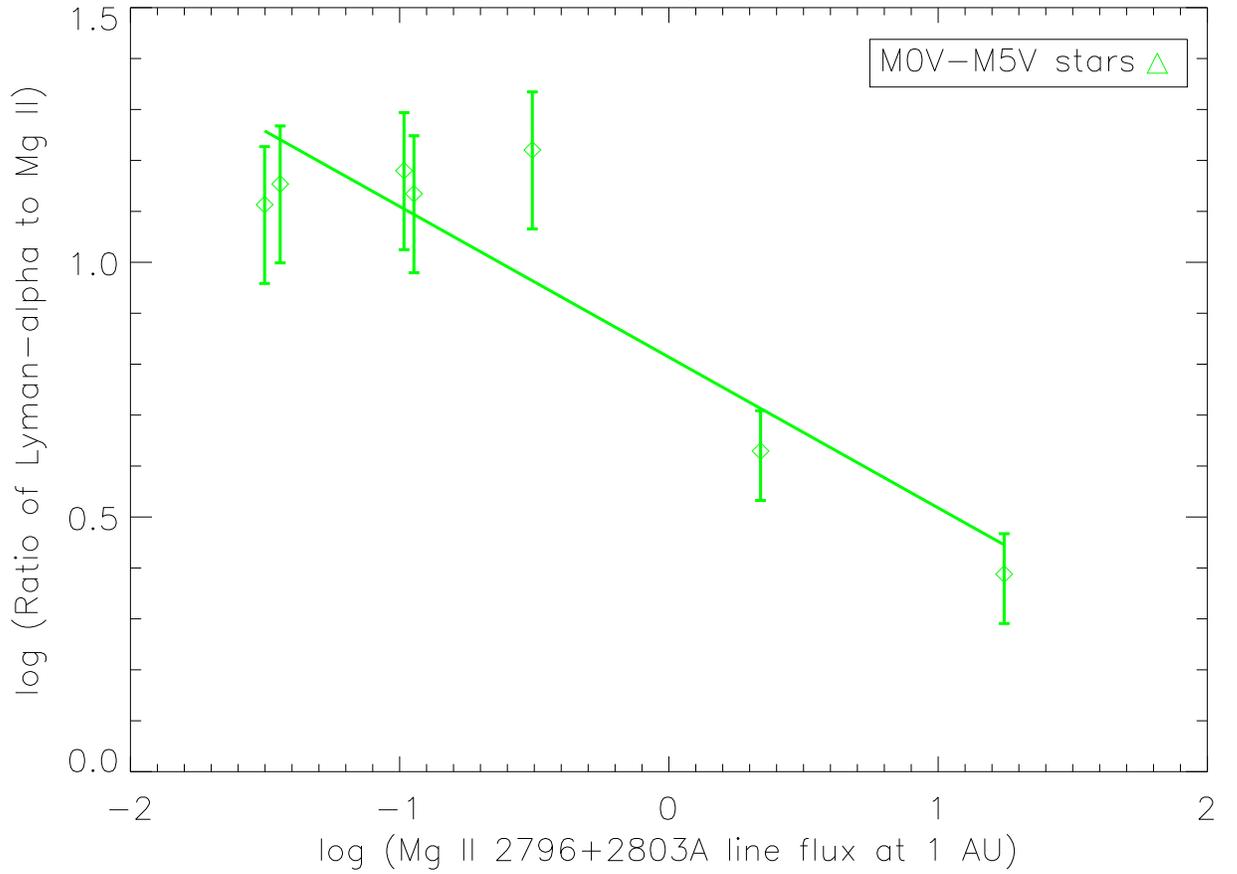}
\caption{Plot of the Lyman-$\alpha$ to Mg~II 2796 + 2803~\AA\ line flux ratio 
vs. the Mg~II line flux at 1 AU for the M0 to M5 stars. The solid line is the 
least-squares fit to the data.  The errors bars are 20\% for stars using the
\citet{Wood2005} correction for the missing Lyman-$\alpha$ flux or 30\% for 
stars using the \citet{France2012a} correction for the missing 
Lyman-$\alpha$ flux.}
\end{figure}

\begin{figure}
\includegraphics{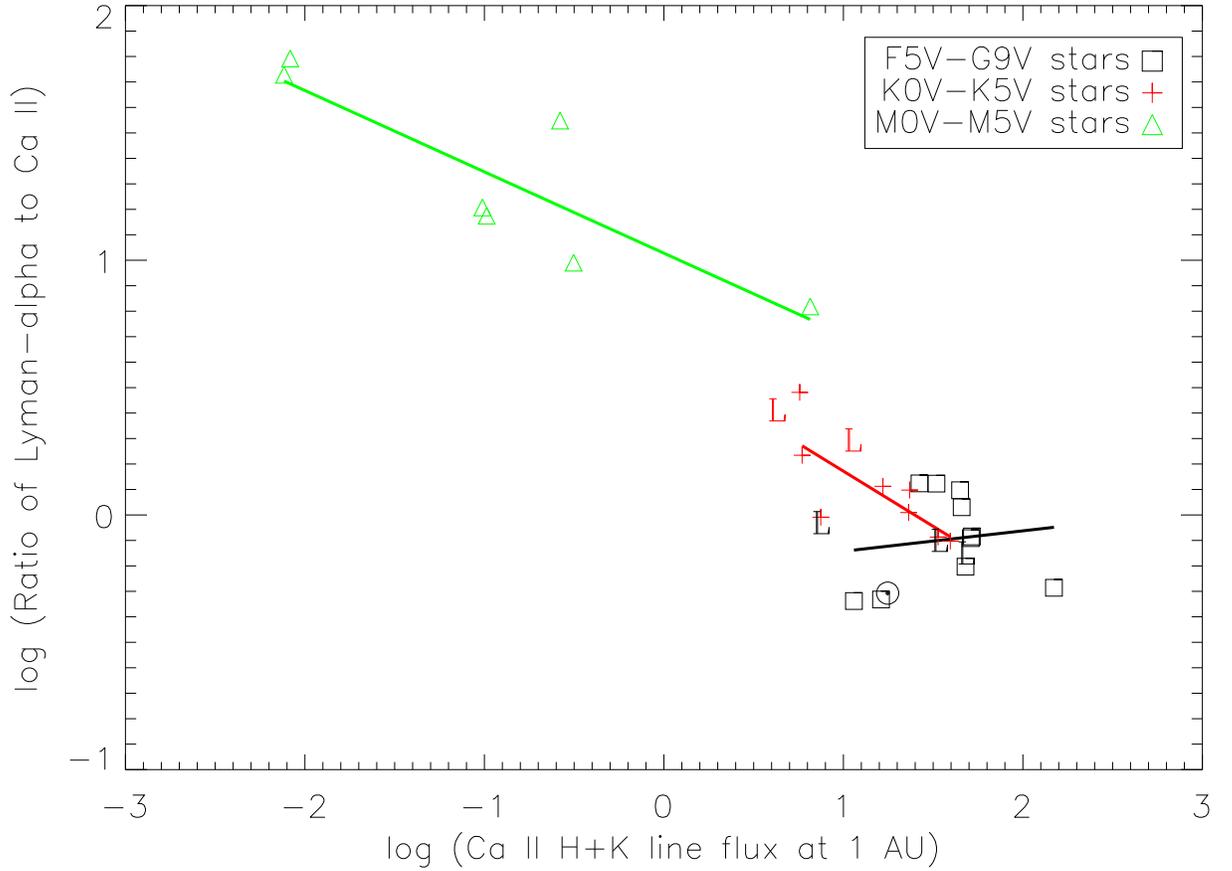}
\caption{Plot of the Lyman-$\alpha$ to Ca~II 3933 + 3968~\AA\ line flux ratio 
vs. the Ca~II line flux at 1 AU for the M0 to M5 stars. 
The solar data are indicated by the Sun symbol,
and the L symbol refers to a star with low metal abundance [Fe/H] $<-0.30$. 
The solid lines are least-squares fits for each spectral type bin
excluding the L stars and the Sun.}
\end{figure}

\begin{figure}
\includegraphics{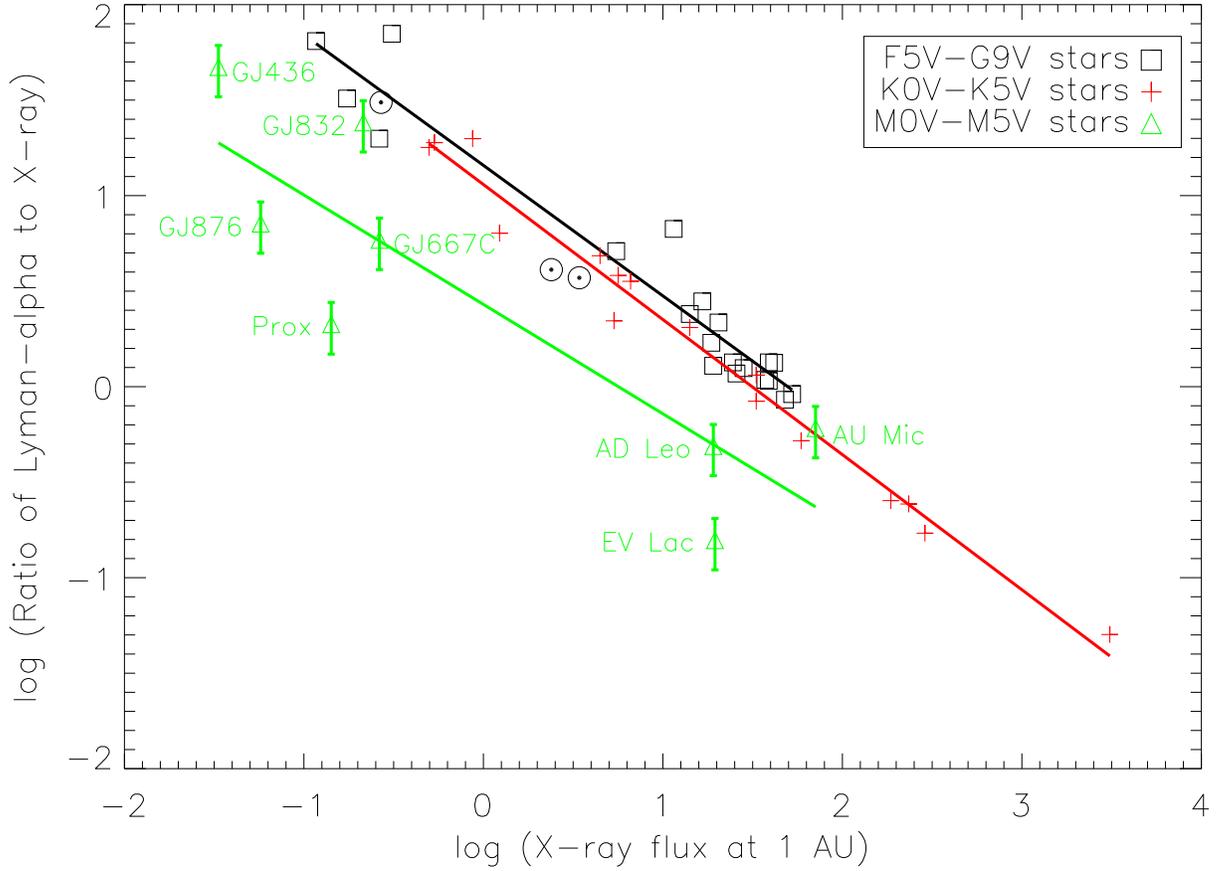}
\caption{Plot of the ratio of Lyman-$\alpha$ to X-ray flux vs. the 
Lyman-$\alpha$ line flux at 1~AU for all stars in our sample with X-ray data.
The solid lines are least-squares fits to the data for the F5~V--G9~V,
K0~V--K5~V, and M0~V--M5~V stars. Data for the quiet Sun, moderately active
Sun, and active Sun are indicated by the dotted circle symbols (from left
to right respectively).M dwarfs stars are identified by name.}
\end{figure}

\begin{figure}
\includegraphics{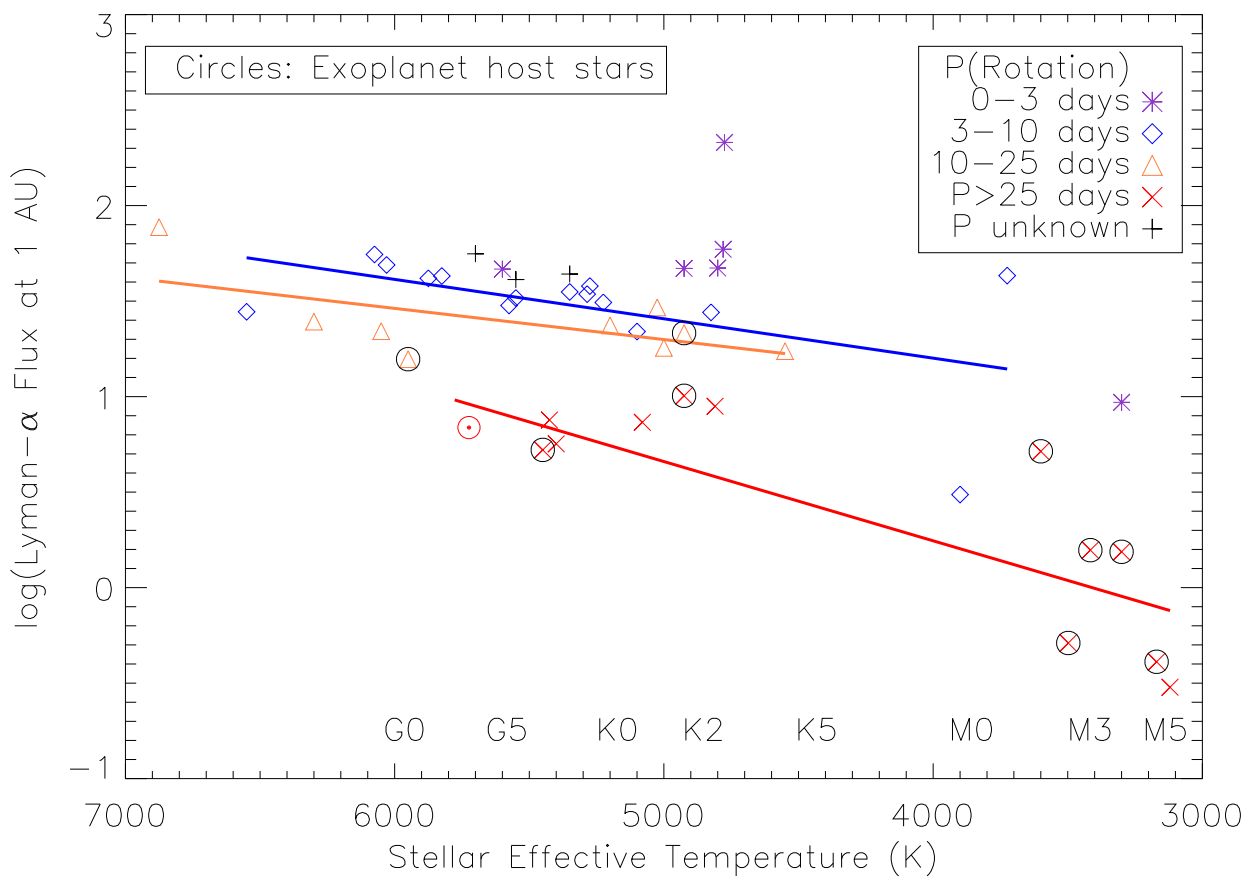}
\caption{Lyman-$\alpha$ flux at 1 AU vs. stellar effective temperature. The 
stars are grouped according to stellar rotation period: ultrafast rotators 
($P_{\rm rot} < 3$ days), fast rotators (3--10 days), moderate rotators
(10--25 days), and slow rotators ($> 25$ days). Rotation period is a rough
measure of the magnetic heating rate in the star's chromosphere and corona.
Host stars of exoplanets are circled and the quiet Sun is marked as a 
circled dot. Least-squares
fit lines are shown for the fast, moderate, and slow rotators.}
\end{figure}

\begin{figure}
\includegraphics{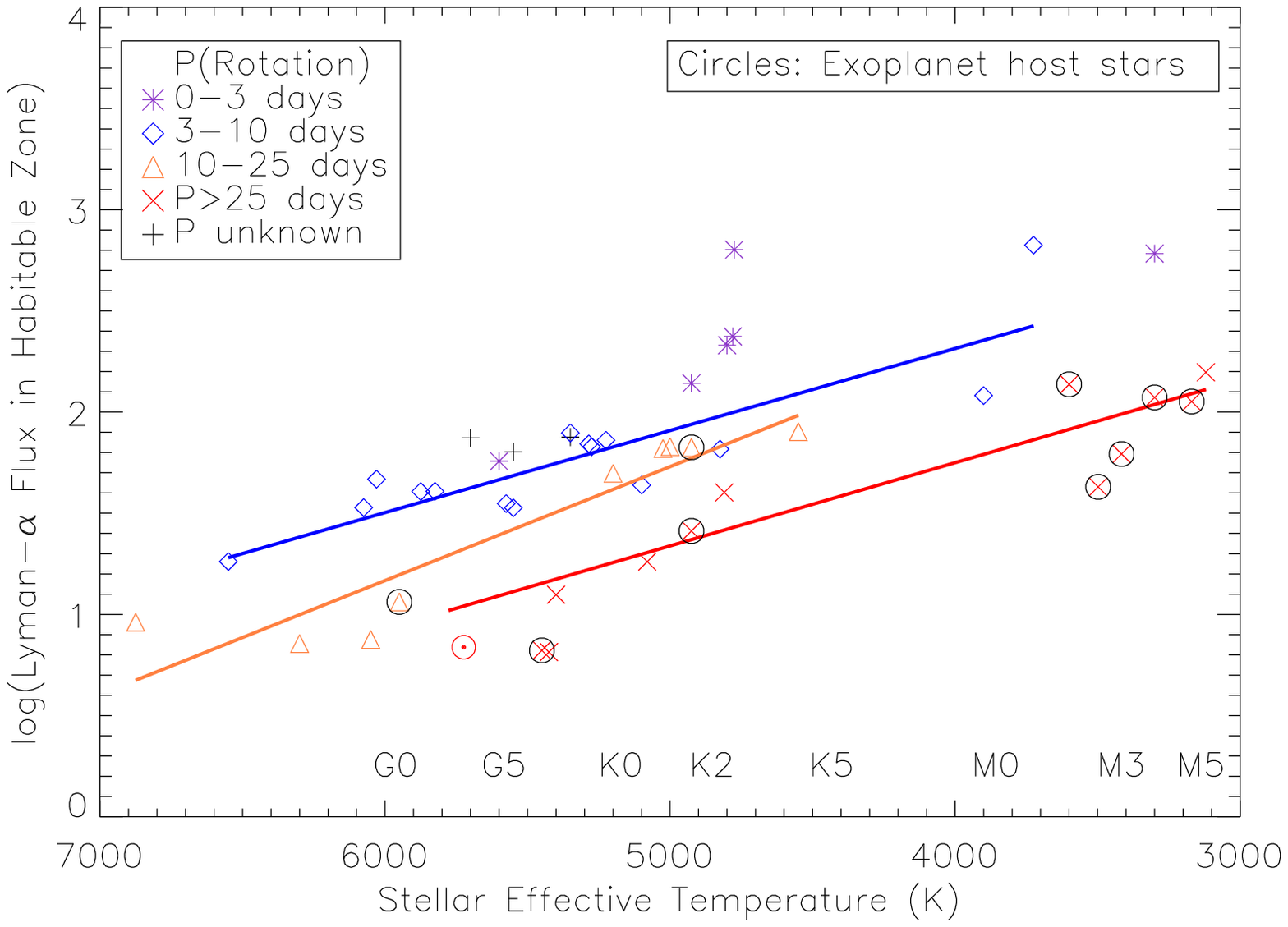}
\caption{Same as Figure 8, except that the Lyman-$\alpha$ flux is evaluated 
in the habitable zone for an exoplanet.}
\end{figure}

\acknowledgements

This work is supported by NASA through grants NNX08AC146, NAS5-98043, 
and HST-GO-11687.01-A to the University of Colorado at Boulder. 
We thank Tom Woods
for providing the SORCE data and Steven Osterman for information on the 
COS calibration.

{\it Facilities:} \facility{HST (COS)}, \facility{HST (STIS)}. 
\facility{SIMBAD}

\end{document}